\documentclass[%
 pop,
 amsmath,amssymb,			
 reprint,%
]{revtex4-1}
\usepackage[utf8]{inputenc}
\usepackage{graphicx}
\usepackage{dcolumn}
\usepackage{bm}
\usepackage{hyperref}
\usepackage{amsmath}
\usepackage{subcaption}
\hypersetup{colorlinks,colorlinks,linkcolor=cyan,anchorcolor=cyan,citecolor=cyan,filecolor=cyan,urlcolor=cyan}
\usepackage[utf8]{inputenc}
\usepackage[T1]{fontenc}
\usepackage{mathptmx}
\usepackage{etoolbox}
\usepackage{lmodern}
\usepackage[capitalize]{cleveref}
\newcommand{\figref}[1]{FIG.\ref{#1}}
\RequirePackage{xpatch}
\xpatchcmd\citenum{\NAT@parfalse}{\NAT@partrue}{}{}

\makeatletter
\def\@email#1#2{%
 \endgroup		
 \patchcmd{\titleblock@produce}
  {\frontmatter@RRAPformat}
  {\frontmatter@RRAPformat{\produce@RRAP{*#1\href{mailto:#2}{#2}}}\frontmatter@RRAPformat}
  {}{}
}%
\makeatother
\begin{document}

\preprint{AIP/123-QED}

\title{Optimization of the Compact Stellarator with Simple Coils at finite-beta}

\author{Haorong Qiu$^1$}\author{Guodong Yu$^2$}\author{Peiyou Jiang$^3$}\author{Guoyong Fu$^1$}
\thanks{corresponding author's Email: gyfu@zju.edu.cn}

\affiliation{$^1$Institute for Fusion Theory and Simulation,School of Physics, Zhejiang University, Hangzhou, China}
\affiliation{$^2$School of Nuclear Science and Technology, University of Science and Technology of China, Hefei, China}
\affiliation{$^3$Max Planck Institute for Plasma Physics, Garching, Germany}

\date{\today}

\begin{abstract}
	An optimized stellarator at finite plasma beta is realized by single-stage optimization of simply modifying the coil currents 
of the Compact Stellarator with Simple Coils (CSSC)[Yu et al., J. Plasma Physics 88,905880306 (2022)]. The CSSC is an optimized stellarator obtained by direct optimization via 
coil shapes, with its coil topology similar to that of the Columbia Non-neutral Torus (CNT) [Pederson et al., Phys. Rev. Lett. 88, 205002 (2002)]. Due to its vacuum-based optimization, the CSSC exhibits detrimental finite beta effects on neoclassical confinement. The results of optimization show that the finite beta effects can be largely mitigated by reducing the coil currents of CSSC.
\end{abstract}

\maketitle	
\section{Introduction}

Tokamaks have long been the leading candidate for magnetic confinement fusion due to their axisymmetric geometry and inherently low levels of neoclassical transport. 
However, the commissioning of the advanced stellarator Wendelstein 7-X (W7-X)\cite{dinklage2018magnetic} in 2015 marked a renaissance for stellarators within the global fusion research community. 
The success of W7-X\cite{beidler2021demonstration} has demonstrated that optimized stellarators with complex three-dimensional coils 
and large-scale engineering can be constructed with the precision required to produce well-defined flux surfaces, excellent neoclassical confinement, and other desirable properties as designed.

Design of advanced stellarators\cite{wobig1999theory} like W7-X employs a two-stage optimization approach to achieve enhanced neoclassical confinement and magnetohydrodynamic stability. 
In the first stage, the plasma equilibrium is optimized to meet specific physics goals, such as reduced transport and improved stability. 
In the second stage, three-dimensional coils are designed to accurately reproduce the desired equilibrium, 
ensuring that the theoretical optimization translates into practical engineering solutions. 
However, the design and fabrication of optimized three-dimensional coils often present significant engineering challenges, 
resulting in high complexity and resource-intensive manufacturing processes. 
Therefore, it is critical to explore alternative stellarator designs with simplified coils that maintain high performance while reducing engineering challenges. 
In response, the Zhejiang University Compact Stellarator (ZCS)\cite{yu2021neoclassically} and the Compact Stellarator with Simple Coils (CSSC)\cite{yu2022existence} 
were developed through direct optimization of coil shapes. 
   
\begin{figure}[!t]
	\centering
	\includegraphics[width=0.4\textwidth, keepaspectratio]{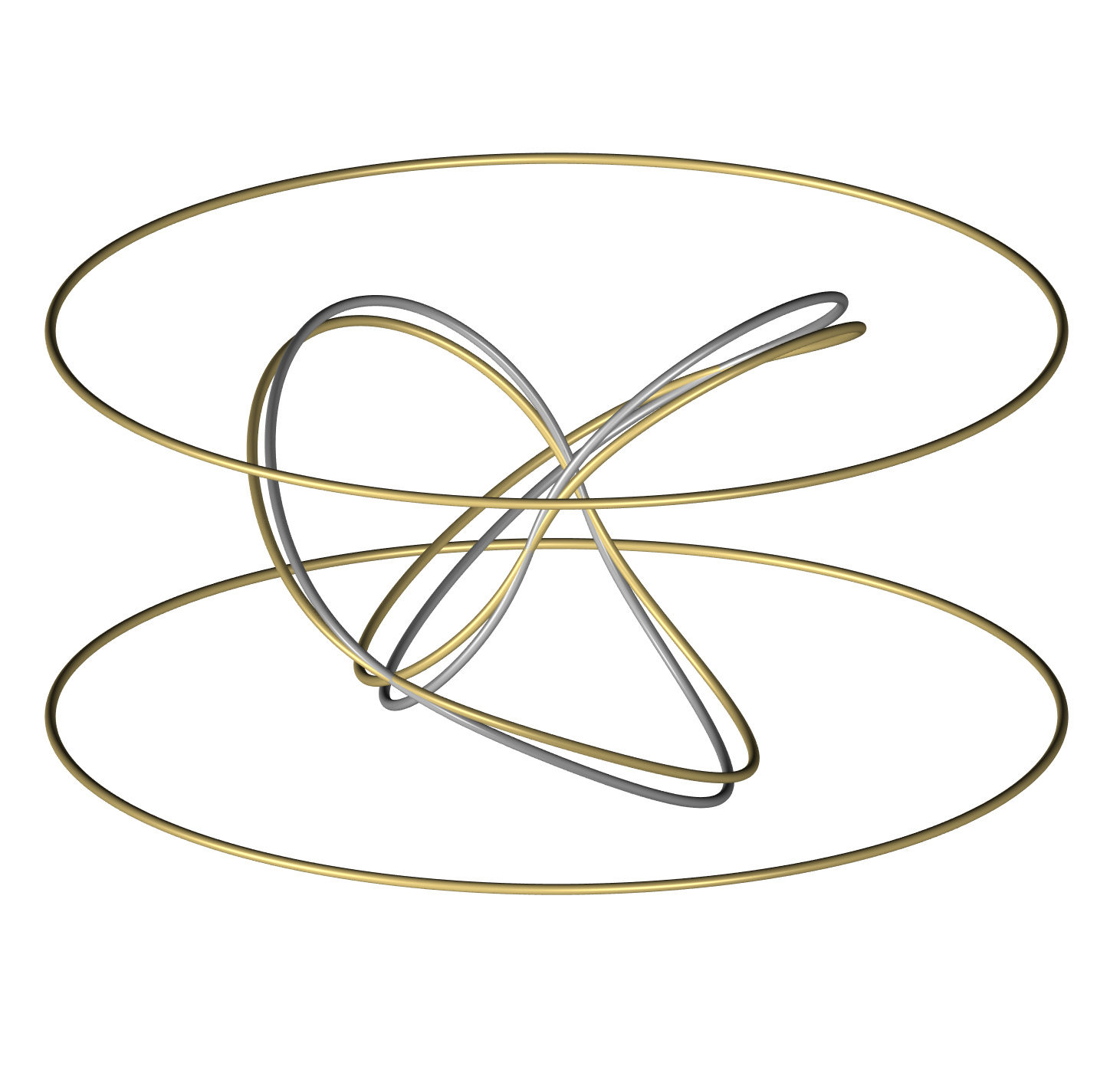} 	
	\caption{The CSSC coil set (gold) and the rotated IL coils of CSSC (silver).}
	\label{fig1}
\end{figure}

\begin{figure}[!t]
	\centering
	\includegraphics[width=0.4\textwidth, keepaspectratio]{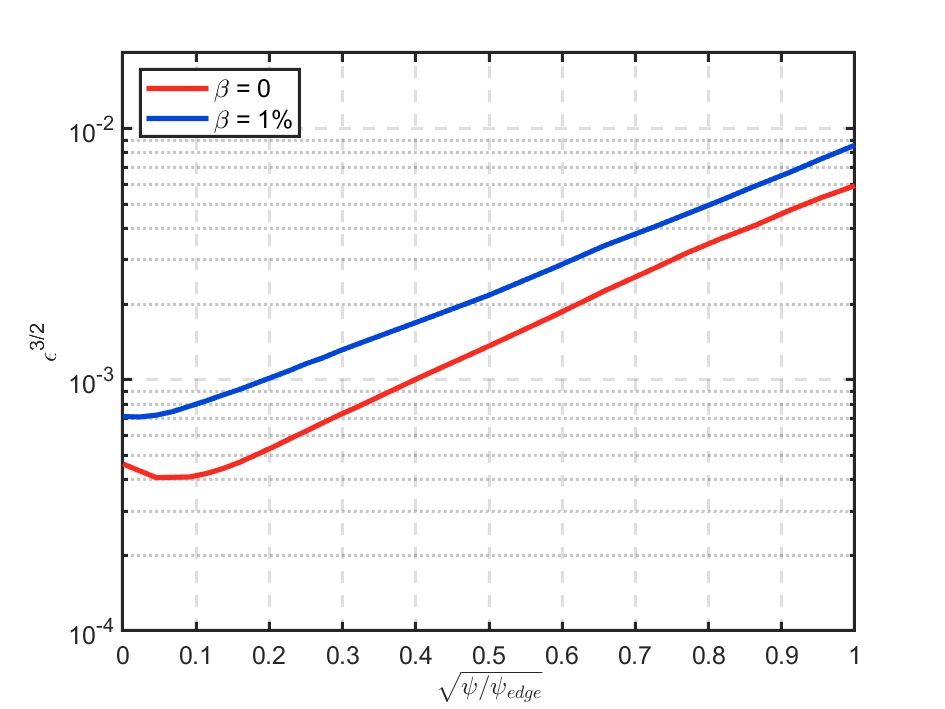} 	
	\caption{Comparison of effective ripple profiles of CSSC for two values of the volume-averaged plasma beta: $\beta=0$ and $\beta=1\%$.}
	\label{fig2}
\end{figure}

The CSSC was obtained through direct variation of coil shapes to simultaneously optimizing plasma confinement and magnetohydrodynamic (MHD) stability. 
The optimization targets for CSSC include key physics parameters such as the effective helical ripple $\epsilon_{eff}^{3/2}$\cite{nemov1999evaluation}, magnetic well depth, plasma volume, rotational transform, and flux surface quality. 
Notably, the effective helical ripple is a critical parameter due to its strong correlation with neoclassical transport. 
The optimized coil configuration is based on the topology of the Columbia Non-neutral Torus (CNT)\cite{pedersen2002confinement}, which consists of two circular interlocking (IL) coils and two circular vertical field (VF) coils. 
\figref{fig1} illustrates the three-dimensional schematic of the CSSC coil sets (gold color).  
The primary difference between the coils of CSSC and CNT is that the IL coils of CSSC are three-dimensional, while those of CNT are circular. 
As a result, the neoclassical confinement and MHD stability of CSSC are optimized. 
CSSC features both a global magnetic well and a low level of effective helical ripple, comparable to that of W7-X, while maintaining relatively simple coil configuration. 
However, as shown in \figref{fig2}, in corresponding finite-beta equilibria, the effective helical ripple is significantly higher at finite beta, leading to a substantial degradation in neoclassical confinement. 
This effect is detrimental to plasma confinement and highlights the critical need for re-optimizing CSSC under finite-beta conditions. 

In this work, we optimize the CSSC configuration to mitigate the finite-beta effect on the effective helical ripple 
using single-stage optimization approach\cite{henneberg2021combined,yu2021neoclassically,yu2022existence} 
by simply varying the currents and orientations of the two inner coils while keeping the coil shape fixed. We will show that the effect of finite beta on the helical ripple can largely be mitigated in this way. Optimized configurations at finite beta values are obtained with low levels of the effective helical ripple comparable to that of the original CSSC design at zero beta.  
The results of this study demonstrate the potential of CSSC for a compact stellarator experimental device with good neoclassical confinement at finite beta. 

The paper is organized as follows: Section II details the optimization methodology, including the parameterization of coil configurations and the computational framework for finite-beta equilibria. 
Section III presents the optimization results for the CSSC at finite beta. 
Finally, Section IV summarizes the key findings, discusses their implications for stellarator design, 
and outlines potential directions for future research. 

\section{Optimization methods}

The optimization procedure comprises three sequential computational steps:
(1) systematic generation of coil configurations through parametric modification of the coil current and coil geometry, 
(2) numerical computation of finite-beta equilibria and 
(3) quantitative evaluation of effective helical ripple based on the computed equilibrium solutions. 

   The coil geometries are mathematically represented through Fourier decomposition in Cartesian space. 
Specifically, each coil in a given configuration is modeled as a continuous, differentiable space curve, 
whose three-dimensional coordinates are parameterized by a truncated Fourier series\cite{zhu2017new}:
\[
\begin{cases}
x=x_{\text{c,0}}+\displaystyle\sum_{n=1}^{n_f}x_{\text{c,n}}cos(nt)\\
\\
y=y_{\text{c,0}}+\displaystyle\sum_{n=1}^{n_f}y_{\text{s,n}}sin(nt)\\
\\
z=z_{\text{c,0}}+\displaystyle\sum_{n=1}^{n_f}z_{\text{s,n}}sin(nt)
\end{cases}
\] 
The angle parameter $t \in [0,2\pi]$, ensures the closure condition of the coil curves. 
The cutoff harmonic number $n_f$ is a key factor for the geometric complexity of the coil shapes, with higher values enabling finer spatial shapes. 
Specifically, our optimization employs $n_f=3$ for the interlocking (IL) coils of the CSSC configuration to accommodate their intricate geometries, 
while a lower harmonic number of $n_f=1$ is sufficient for the simpler VF coils. 
The complete sets of Fourier coefficients for both IL and VF coils are provided in Yu's paper\cite{yu2022existence} for the CSSC configuration. 
Exploiting the intrinsic symmetry properties of the coil system, the Cartesian coordinates of the second IL coil and the second VF coil can be derived through the following transformations: 
\[
\begin{aligned}
\begin{cases}
x_{_{IL2}}=-x_{_{IL1}}\\

y_{_{IL2}}=-y_{_{IL1}}\\

z_{_{IL2}}=z_{_{IL1}}
\end{cases}
\end{aligned}
\begin{aligned}
\begin{cases}
x_{_{VF2}}=x_{_{VF1}}\\

y_{_{VF2}}=y_{_{VF1}}\\

z_{_{VF2}}=-z_{_{VF1}}
\end{cases}
\end{aligned}
\]
where the index IL1 and VF1 refer to the first IL coil and the first VF coil respectively, and IL2 and VF2 refer to the second IL coil and VF coil. The current ratio of the CSSC's IL coils and VF coils is $I_{IL}/I_{VF}=1.323$. 
This specific current ratio is critical for achieving the optimized properties of CSSC. 

The optimization degrees of freedom comprise of three key parameters: (1) the rotation angle ($\delta \theta$) of the IL coils about the x-axis with respect to the original IL coils, 
(2) the normalized IL coil current $\overline{I}=I_{IL}/I_{VF}$, and (3) the vertical displacement ($\delta h$) of the VF coils. 
Here, $I_{VF}$ remains fixed at its nominal value, while $\overline{I}$ serves as a dimensionless scaling factor for the IL coil current relative to the VF coil current.
The rotation angle $\delta \theta$ is defined as the angular displacement about the x-axis between the original CSSC IL coils and the new (rotated) IL coils. 
\figref{fig1} illustrates the IL coil configuration (silver color) for $\delta \theta=0.02\pi$, highlighting the geometric transformation due to the rotation. 
The Cartesian coordinate transformation relating the coordinates of the rotated IL coil to the original coordinates is given by the following : 
\[
\begin{cases}
x^{'}(t)=x(t)\\
\\
y^{'}(t)=y(t) cos(\delta \theta)-z(t) sin(\delta \theta)\\
\\
z^{'}(t)=z(t) cos(\delta \theta)+y(t) sin(\delta \theta)
\end{cases}
\]
where $x(t)$, $y(t)$, and $z(t)$ is the Cartesian coordinates of CSSC IL coil, and the $x^{'}(t)$, $y^{'}(t)$, and $z^{'}(t)$ is the Cartesian coordinates of the rotated coils. 
Finally, $\delta h$ is defined as the vertical displacement of the top VF coil and $h_0=0.77m$ is the z-coordinate of CSSC's top VF coil. Thus, the hight of the shifted VF coil is $h=h_0+\delta h$

The optimization targets the minimization of the effective helical ripple $\epsilon_{eff}^{3/2}$, a critical parameter governing neoclassical transport in stellarators. 
This metric is particularly significant for the $1/\nu$ neoclassical transport\cite{kovrizhnykh1984neoclassical, ho1987neoclassical}, where $\nu$ is the plasma collision frequency. 
The $1/\nu$ neoclassical transport presents significant challenges for stellarator confinement in high-temperature regimes, as the associated transport scales inversely with collision frequency and consequently increases dramatically with plasma temperature. 
This strong temperature dependence necessitates rigorous minimization of neoclassical transport in the $1/\nu$ regime for viable stellarator reactors. 
Theoretical and numerical studies have established a direct proportionality between $1/\nu$ transport and the effective helical ripple coefficient $\epsilon_{eff}^{3/2}$\cite{nemov1999evaluation}, 
making $\epsilon_{eff}^{3/2}$ a critical optimization parameter for confinement improvement. 
Theoretical analyses have established that $\epsilon_{eff}^{3/2}$ is uniquely determined by the magnetic field geometry and can be computed through field-line integration\cite{nemov1999evaluation}. 
However, in finite-beta equilibria, the self-consistent magnetic configuration deviates significantly from the vacuum field solution due to plasma pressure effects. 
Consequently, the computational workflow requires sequential steps: calculation of finite-beta equilibria, 
followed by evaluation of the effective helical ripple based on the computed magnetic field structure. 
The following details this computational procedure for evaluation of optimization objectives. 
   
For each coil configuration, the magnetic field is computed through numerical implementation of the Biot-Savart law using a piecewise linear discretization approach\cite{hanson2002compact}.  
The magnetic flux surfaces are then constructed by tracing magnetic field lines with high precision. Based on these flux surfaces, the last closed flux surface (LCFS) are then determined. 
The free-boundary equilibria are computed using the VMEC\cite{hirshman1983steepest}. The input includes the Fourier coefficients of the last closed flux surface (LCFS) calculated above as initial guess, the plasma pressure profile, 
and the bootstrap current profile. For free-boundary calculations, the input also includes the vacuum magnetic field from coils on a 3D grid. 
Subsequently, the effective helical ripple is evaluated using the NEO code based on the computed finite beta equilibrium. 

The pressure profile is prescribed as $p=p_0(1-s^2)^3$, where $p_0$ denotes the on-axis pressure 
and $s=\psi/\psi_{edge}$ represents the normalized toroidal flux. 
A fixed bootstrap current profile, obtained from the self-consistent CSSC equilibrium at the volume-averaged plasma beta of $\beta=1\%$, is utilized as the reference profile for optimization. 
In principle, during optimization, the bootstrap current profile varies due to changes in magnetic configuration. However, we will show below that this non-self-consistency affects little the optimization targets.

The bootstrap current profile is computed using SFINCS, 
a kinetic code that solves the steady-state drift-kinetic equation and bootstrap current for multiple species\cite{landreman2014comparison}. 
Achieving a converged bootstrap current profile typically requires multiple iterations between free-boundary equilibrium calculations and bootstrap current computations using SFINCS. 
However, the computational cost of maintaining self-consistency during the optimization process would be prohibitively high, 
as each coil configuration would necessitate this iterative procedure. 
Therefore, during the optimization process, a fixed bootstrap current profile is employed for computational efficiency. 
While this results in non-self-consistent equilibria, the impact on the optimization target is assumed to be negligible. 
Once an optimized configuration is obtained, a fully self-consistent bootstrap current and equilibrium are recomputed. As demonstrated in the following section, 
the effect of this approximation on the effective helical ripple, our primary optimization target, is minimal, validating the initial assumption. 

Although a fixed bootstrap current profile is employed during optimization, 
the computational cost per each configuration is still significantly higher compared to that of vacuum field calculations. 
This computational challenge imposes substantial constraints on the degrees of freedom available for optimization for finite-beta equilibria. 
Therefore, we first explore optimization using only $\delta \theta$ and $\overline{I}$  as the primary degrees of freedom. 
Subsequently, the influence of the additional parameter of $\delta h$ is investigated to assess its impact on the optimization target. As will be shown below, these limited degrees of freedom are mostly sufficient for achieving our goal of optimization. 

\section{Optimization results}
\begin{figure}[!t]
	\centering
	\includegraphics[width=0.4\textwidth, keepaspectratio]{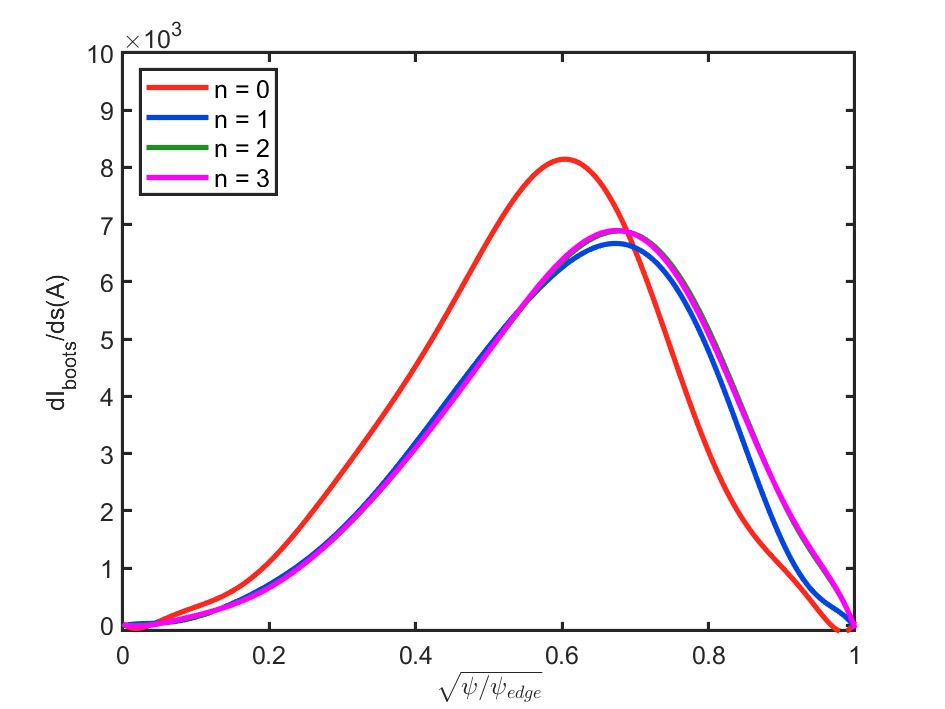} 	
	\caption{Comparison of bootstrapping current profiles from iteration process. Different colors correspond to different iteration steps.}
	\label{fig3}
\end{figure}

\begin{figure}[!t]
	\centering
	\includegraphics[width=0.4\textwidth, keepaspectratio]{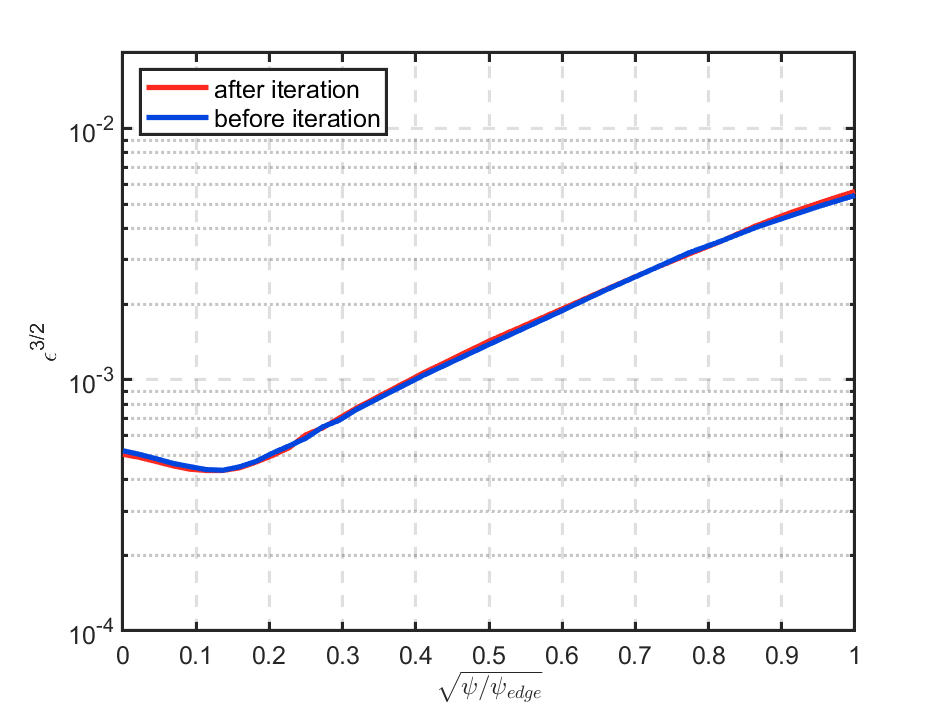} 	
	\caption{The profile of the effective helical ripple $\epsilonˆ{3/2}$ before and after iteration.}
	\label{fig4}
\end{figure}
Using the optimization framework described above, we obtain an optimized coil configuration and corresponding equilibrium at each value of beta. 
While the initial equilibrium is non-self-consistent, self-consistency is achieved through iteration of the bootstrap current. 
\figref{fig3} demonstrates the convergence behavior of the bootstrap current iterations, showing near-complete convergence within only three iterations, 
where $\psi_{edge}$ denotes the poloidal flux at plasma edge. 
In the figure, the curve labeled $n=0$ represents the fixed bootstrap current profile used during the optimization process. 
When computing the bootstrap current using SFINCS, a uniform temperature profile is assumed for simplicity, 
while the density profile is consistent with the pressure profile used in the VMEC calculations. 
The peak electron and ion densities are set to be $n_{e0}=n_{i0}=1.2\times10^{19}m^{-3}$, 
with electron and ion temperatures of $T_{e0}=T_{i0}=100eV$ at volume-averaged beta of $\beta=1\%$. 
\figref{fig4} compares the profiles of the effective helical ripple for two equilibria: the optimized equilibrium with the fixed bootstrap current profile 
and the one with the converged bootstrap current profile. 
The minimal difference between these two profiles demonstrates that the impact on optimization target of the effective helical ripple is negligible, 
thereby validating the initial procedure of using a fixed bootstrap current profile during optimization. 

In this section, we prioritize the optimization of the IL coils due to their dominating impact on the magnetic configuration. 
The optimization is performed by varying $\delta \theta$ and $\overline{I}$ at $\beta=1\%$, 
with parameter ranges of $[-0.02\pi,0.02\pi]$ for $\delta \theta$ and $[1.1,1.5]$ for $\overline{I}$. 
A straightforward grid search method is employed, with grid sizes of $0.001\pi$ for $\delta \theta$ and $0.001$ for $\overline{I}$, ensuring thorough exploration of the 2D parameter space. 
This section is organized as follows: Subsection A analyzes the influence of $\overline{I}$ and $\delta \theta$ on the effective helical ripple, demonstrating that a moderate reduction in $\overline{I}$ largely mitigates finite-beta effects. 
Subsection B details the optimized coil configuration and compares its targets for different beta values. 
Subsection C further refines the configuration by adjusting the parameter $\delta h$ of the vertical field coil, 
reducing the parameter $\delta h$ exhibits a similar effect on the effective helical ripple as decreasing $\overline{I}$. 
Subsection D evaluates the MHD stability of the optimized configurations at finite beta. 
Finally, Subsection E investigates the critical beta for global MHD stability. 

\subsection{The effect of varying $\overline{I}$ and $\delta \theta$} 

\begin{figure}[!t]
	\centering
	\includegraphics[width=0.4\textwidth, keepaspectratio]{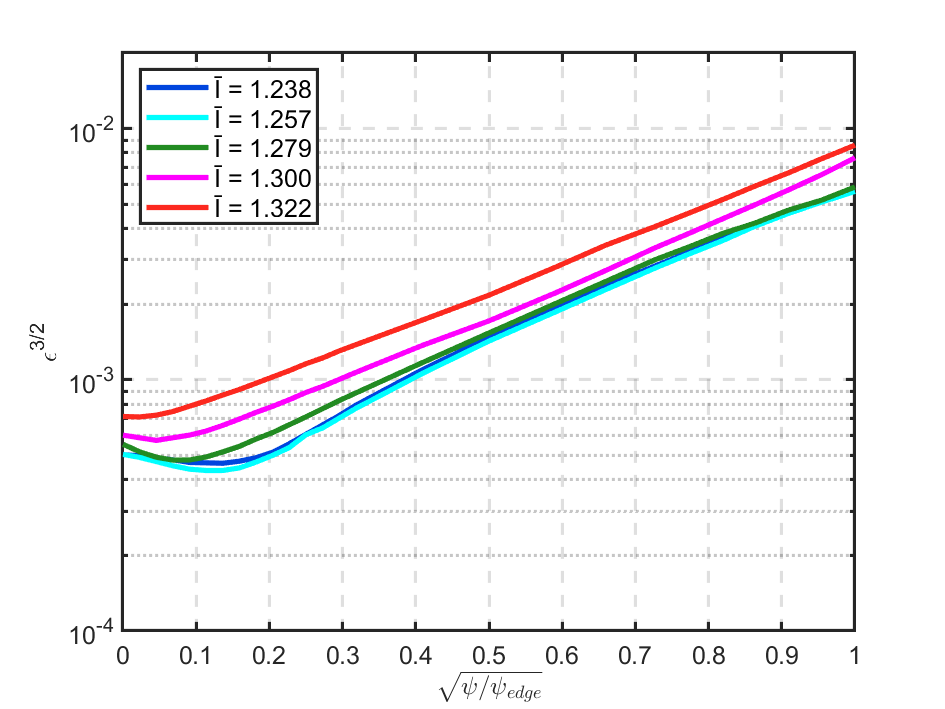} 	
	\caption{The profiles of effective helical ripple of different $\overline{I}$ and a fixed value of $\delta \theta=0$.}
	\label{fig5}
\end{figure}

\begin{figure}[!t]
	\centering
	\includegraphics[width=0.4\textwidth, keepaspectratio]{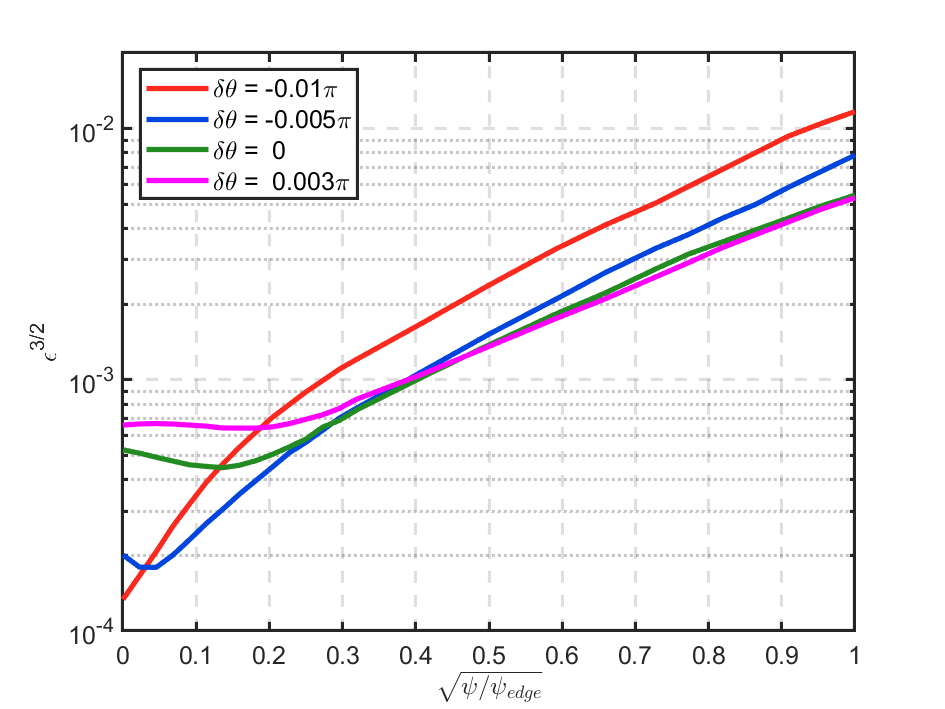} 	
	\caption{The profiles of effective helical ripple for different $\delta \theta$.}
	\label{fig6}
\end{figure}

\figref{fig5} presents the change of the effective helical ripple profile for varying $\overline{I}$ value at $\delta \theta=0$ and $\beta = 1\%$. 
As $\overline{I}$ decreases, the overall effective ripple is reduced, reaching a minimum at $\overline{I}\approx1.257$. This optimal value represents a $5\%$ reduction from the original value of $\overline{I}=1.323$.
Further reduction of $\overline{I}$ beyond this point results in a slight increase in the effective ripple. 

\figref{fig6} illustrates the effective ripple profiles for different values of $\delta \theta$ at a fixed $\overline{I}\approx1.25$ 
(the effective ripple is nearly minimized at $\overline{I}=1.25$ for different values of $\delta \theta$). 
As $\delta \theta$ decreases, the effective helical ripple near the last closed flux surface (LCFS) increases, while the ripple near the magnetic axis decreases. On the other hand, increase of $\delta \theta$ does not change the ripple except near the magnetic axis where the ripple is enhanced. Given the optimization objective of minimizing the effective helical ripple near the outer magnetic surfaces, 
we prioritize parameter adjustments that significantly influence this region. 
This focus ensures that the optimized configuration achieves enhanced confinement properties near the plasma edge, 
which is critical for overall stellarator performance. 
Based on these results, the optimal parameters at $\beta=1\%$ are found to be $\delta \theta=0$ and $\overline{I}=1.258$. 
The results demonstrate that retaining the original value of $\delta \theta$ 
while moderately reducing $\overline{I}$ yields the maximum reduction in effective helical ripple.

The following subsection provides a detailed analysis of this optimized configuration. 

\subsection{The optimized configuration}
\begin{figure}[!t]
	\centering
	\label{fig7}
	\begin{subfigure}[b]{0.4\textwidth}
		\includegraphics[width=\textwidth]{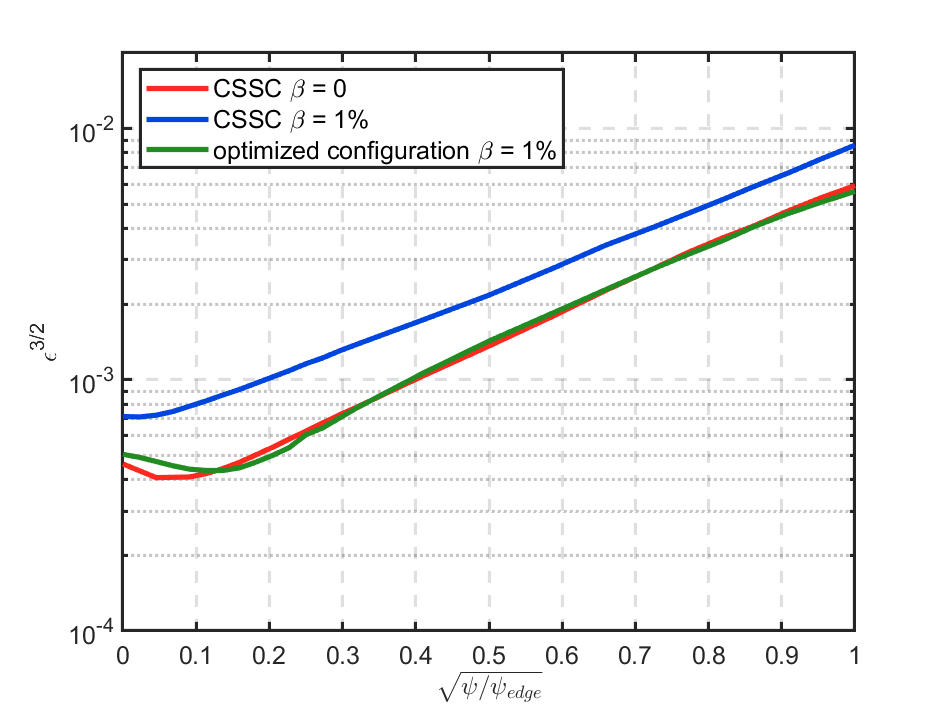} 
		\caption{}
		\label{fig7.a}
	\end{subfigure}
	\quad
	\begin{subfigure}[b]{0.4\textwidth}
		\includegraphics[width=\textwidth]{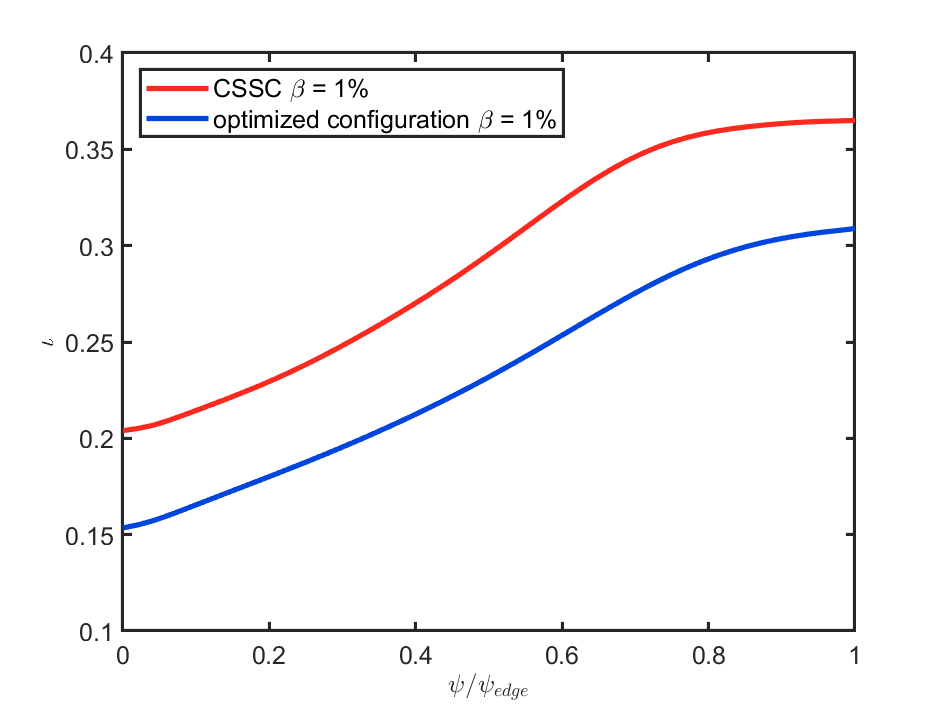} 
		\caption{}
		\label{fig7.b}
	\end{subfigure}
    \caption{Comparison of (a)profiles of the effective ripple $\epsilon^{3/2}$
	and (b)profiles of rotational transform between the optimized configuration 
	and CSSC.}
\end{figure}

\begin{figure}[!t]
	\centering
	\includegraphics[width=0.5\textwidth, keepaspectratio]{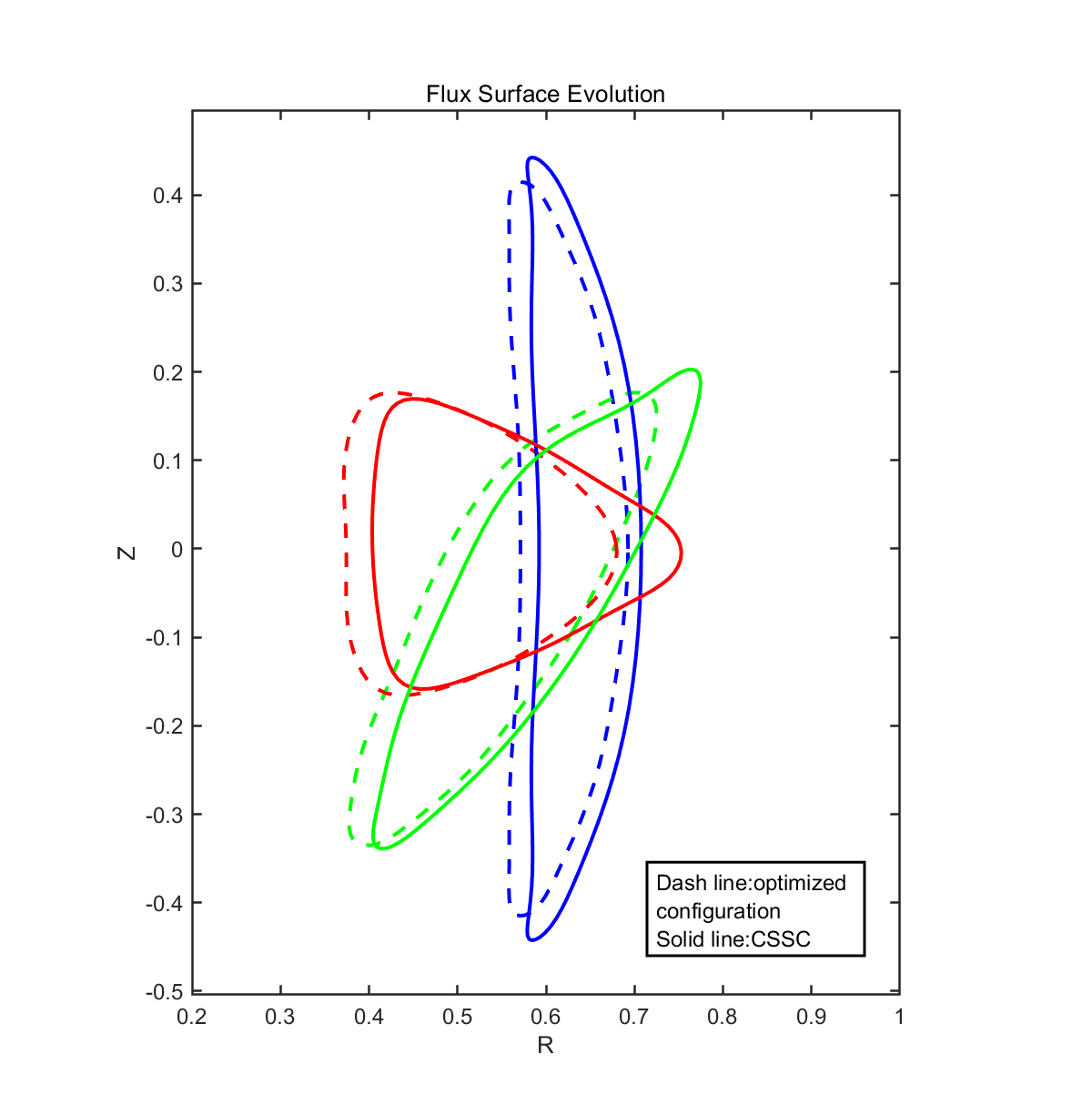} 	
	\caption{Cross sections of the boundary magnetic surface for CSSC (solid lines) and the optimized configuration (dashed lines) at $\beta=1\%$.}
	\label{fig8}
\end{figure}

    \figref{fig7.a} compares profiles of the effective helical ripple $\epsilon^{3/2}$ between CSSC and the optimized configuration at $\beta=1\%$. 
The optimized configuration exhibits a significant reduction in effective helical ripple compared to that of CSSC at $\beta=1\%$, 
achieving a ripple level comparable to that of CSSC in vacuum ($\beta=0$). 
This result demonstrates the success of our optimization in reducing the effective helical ripple at finite beta by only varying coil current. 

Furthermore, \figref{fig7.b} presents the rotational transform profiles for both configurations at $\beta=1\%$, 
revealing that the rotational transform of the optimized finite-beta equilibrium is slightly lower than that of CSSC. 
Finally, \figref{fig8} compares the cross-sections of the boundary magnetic surfaces of CSSC (solid lines) and the optimized configuration (dashed lines) at $\beta=1\%$. 
The optimized configuration exhibits an inward shift in the boundary magnetic surface compared to that of CSSC, 
resulting in a slightly more compact configuration. 
\begin{figure}[!t]
	\centering
	\includegraphics[width=0.4\textwidth, keepaspectratio]{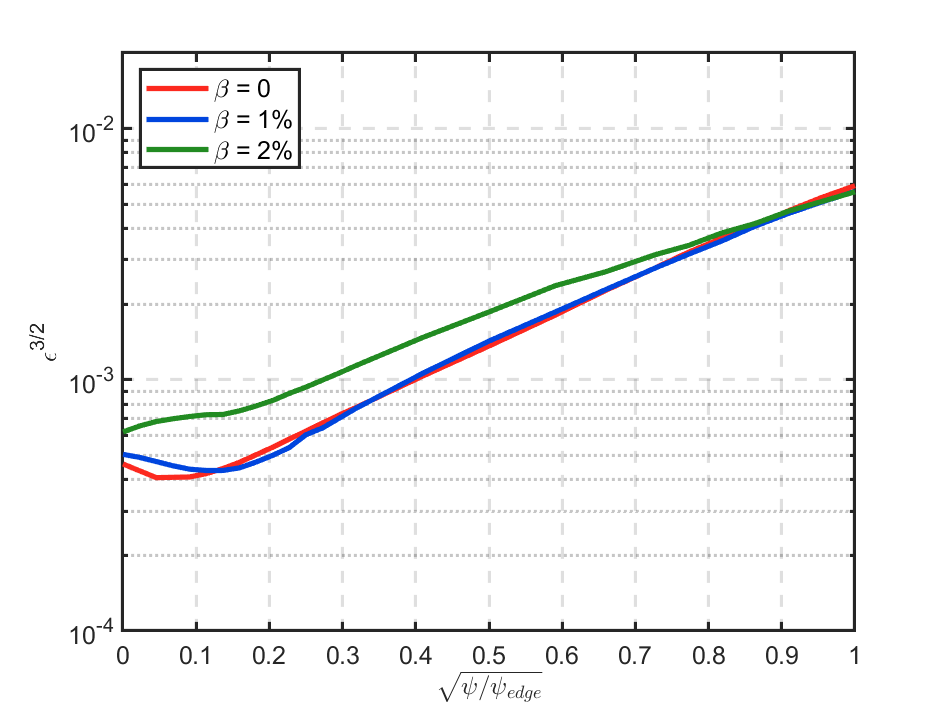} 	
	\caption{Effective ripple profiles of optimized configuration at different beta values.}
	\label{fig9}
\end{figure}

Next, we compare the ripple levels of the optimized configurations at different beta values. 

For higher beta values, we maintain a constant temperature profile 
($T_{e0}=T_{i0}=100eV$) while scaling the density. For $\beta\approx1\%$, the peak electron and ion densities are $n_{e0}=n_{i0}=1.2\times10^{19}m^{-3}$, 
while for $\beta\approx2\%$, the densities are increased to $n_{e0}=n_{i0}=2.4\times10^{19}m^{-3}$. 
\figref{fig9} compares the effective helical ripple profiles of optimized configurations at $\beta=0\%$, $\beta=1\%$ and $\beta=2\%$. 
We observe that the finite-beta effects on the effective helical ripple are nearly eliminated at $\beta=1\%$ by optimization. On the other hand, for the $\beta=2\%$ case, the optimization is still successful in reducing the ripple level near the plasma edge. However, in the inner region, the ripple level is increased. Thus, the mitigation of the finite beta effects is not complete at this beta value for the inner part of plasma. It should be noted that the optimal coil parameters for this case are $\delta \theta=0.003 \pi$, $\overline{I}=1.25$. 


\subsection{The effect of changing $\delta h$}
\begin{figure}[!t]
	\centering
	\includegraphics[width=0.4\textwidth]{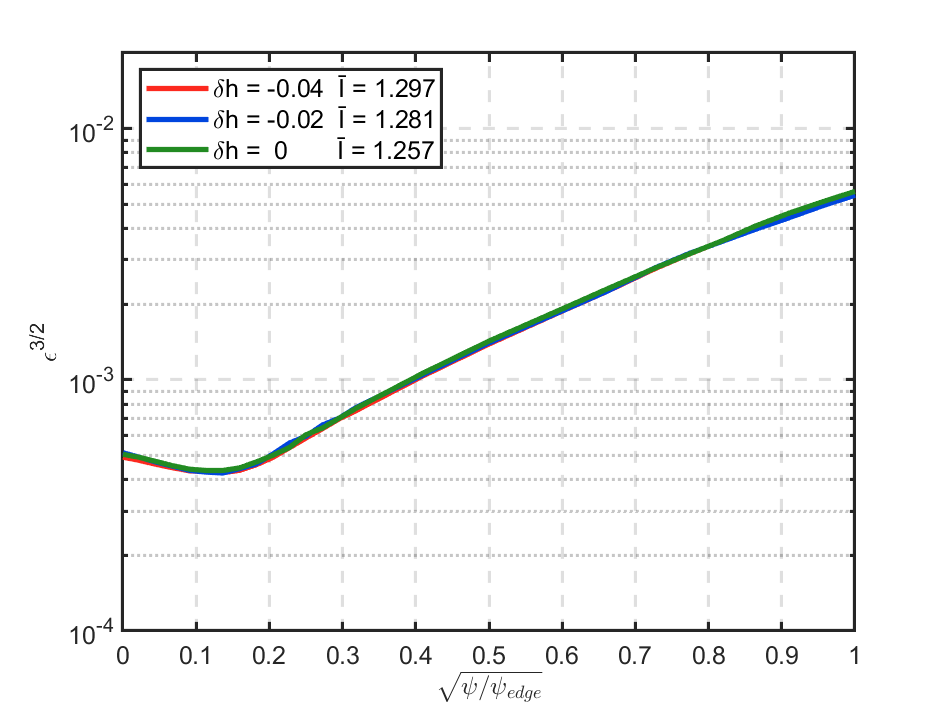} 	
	\caption{The effective ripple profiles of optimized configuration with different $\delta h$.}
	\label{fig10}
\end{figure}
    We further explore additional degrees of freedom, specifically the vertical displacement parameter $\delta h$ of the vertical field coil, 
to optimize the effective helical ripple profile at $\beta=1\%$. 
Initially, we vary only $\delta h$ with the original values of $\delta \theta=0$ and $\overline{I}=1.323$ and obtain an optimal value of $\delta h\approx -0.05 m$, which significantly reduces the effective helical ripple. 
At $\delta h\approx -0.05 m$, the effective helical ripple is optimized to a level comparable to that achieved through varying $\overline{I}$. 
Next, we simultaneously vary $\delta h$ and $\overline{I}$, revealing an inverse relationship between these two parameters when the effective helical ripple is minimized: 
larger values of $\delta h$ correspond to smaller values of $\overline{I}$. 
Furthermore, the optimized effective helical ripple profiles for different $(\delta h,\overline{I})$ combinations are remarkably similar, as shown in \figref{fig10}. This result is understandable since reducing the vertical hight of the vertical field coils is nearly equivalent to increasing the coil current $I_{VF}$ with respect to the current ratio assuming that the vertical magnetic field is approximately uniform in the region of flux surfaces. We note that the vertical field increases as the distance between the two vertical field coils decreases. 

\subsection{MHD stability}
For MHD stability analysis of stellarators, the Mercier criterion\cite{mercier1962stability, greene1962stability, rosenbluth1957stability} is widely employed. 
This criterion is closely related to magnetic well, which has been shown to provide stabilizing effects on interchange modes in toroidal equilibria with isotropic pressures\cite{furth1964closed, johnson1967resistive}. 
Therefore, the magnetic well can serve as a simplified measure of MHD stability. The definition of the magnetic well is given as
\[
   W=2\frac{V}{\langle{B^2}\rangle}\frac{d}{dV}\langle{\frac{B^2}{2}}\rangle
\]
where $V$ denotes plasma volume within a magnetic surface and the angle brackets represent the magnetic surface average:
\[
	\langle{f}\rangle=\frac{\int\frac{fdl}{B}}{\int\frac{dl}{B}}
\]
Here, $\int dl$ represents the integral along the magnetic field line. 
The above magnetic well formula can be interpreted as the radial gradient of the average magnetic pressure across different magnetic surfaces. 
A positive gradient indicates a magnetic well structure, which is stabilizing, while a negative gradient corresponds to a magnetic hill structure, which is destabilizing.

\begin{figure}[!t]
	\centering
	\includegraphics[width=0.4\textwidth, keepaspectratio]{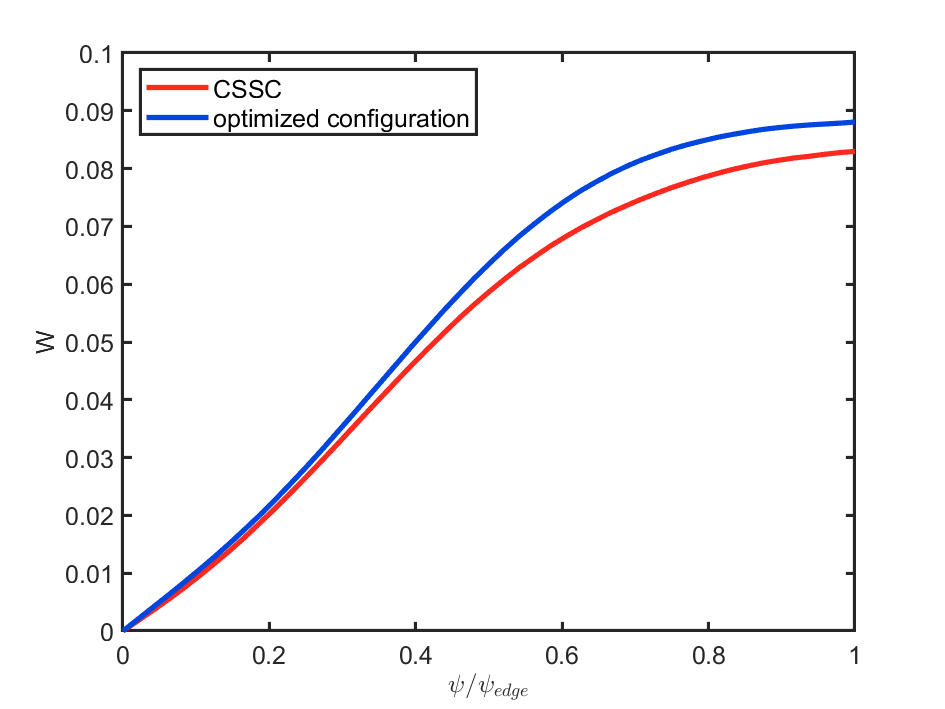} 
    \caption{The comparison of magnetic well profiles between the optimized configuration and CSSC at $\beta=1\%$ equilibria}
	\label{fig11}
\end{figure}

   \figref{fig11} compares the magnetic well profiles between the optimized configuration and the CSSC at $\beta=1\%$. 
As shown in the figure, the optimized configuration exhibits a slight increase in magnetic well compared to that of the CSSC. 
This indicates that the MHD stability is not only preserved but is slightly enhanced by the optimization process. 


\subsection{Critical beta of global MHD stability}
\begin{figure}[!t]
	\centering
	\includegraphics[width=0.4\textwidth, keepaspectratio]{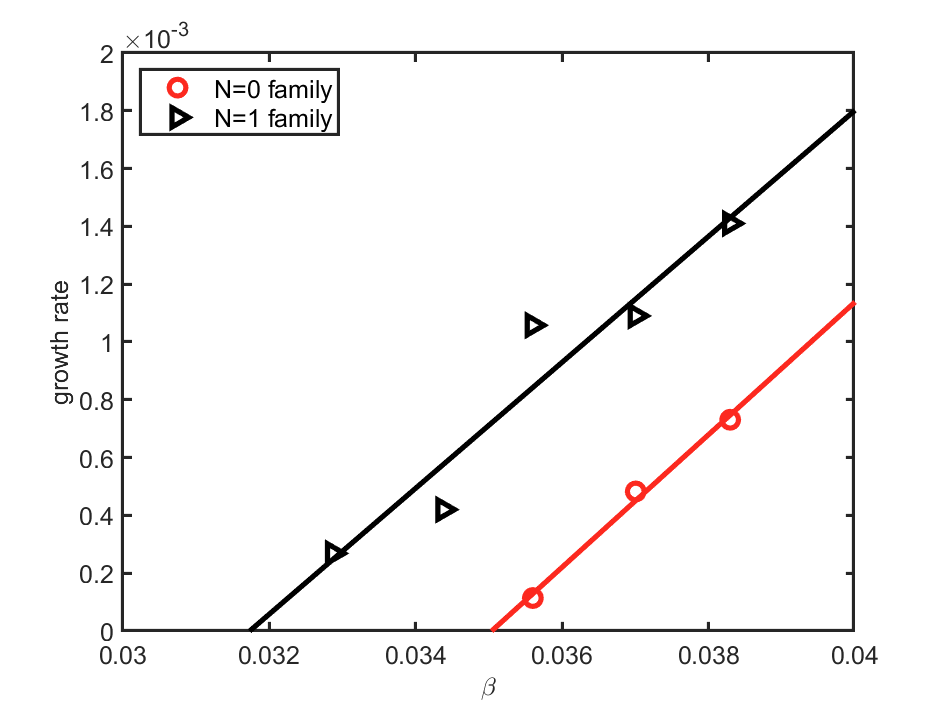} 	
	\caption{The normalized growth rate as a function of plasma beta.}
	\label{fig12}
\end{figure}

\begin{figure}[!t]
	\centering
	\begin{subfigure}[b]{0.4\textwidth}
		\includegraphics[width=\textwidth]{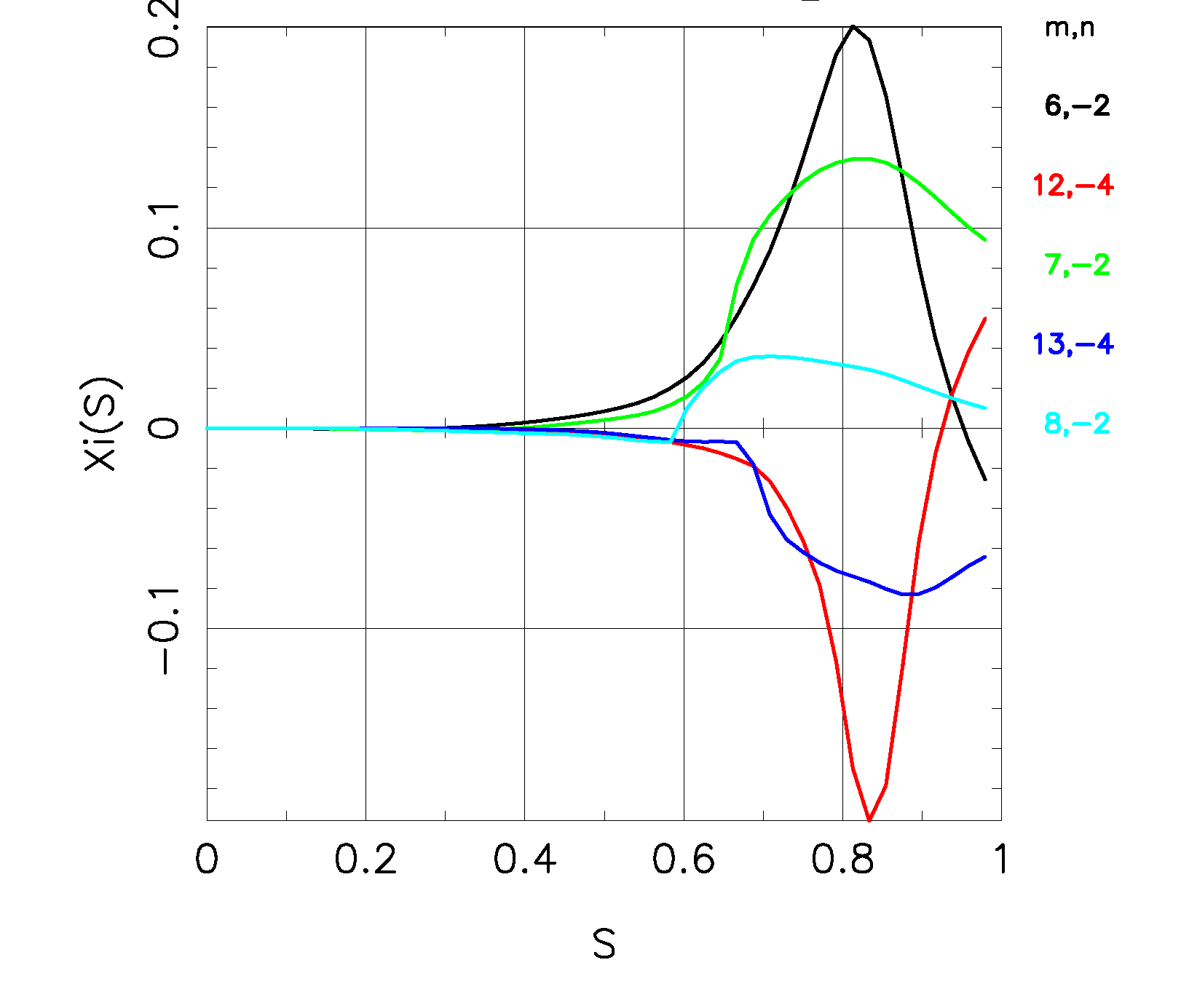} 
		\caption{}
		\label{fig13.a}
	\end{subfigure}
	\quad
	\begin{subfigure}[b]{0.4\textwidth}
		\includegraphics[width=\textwidth]{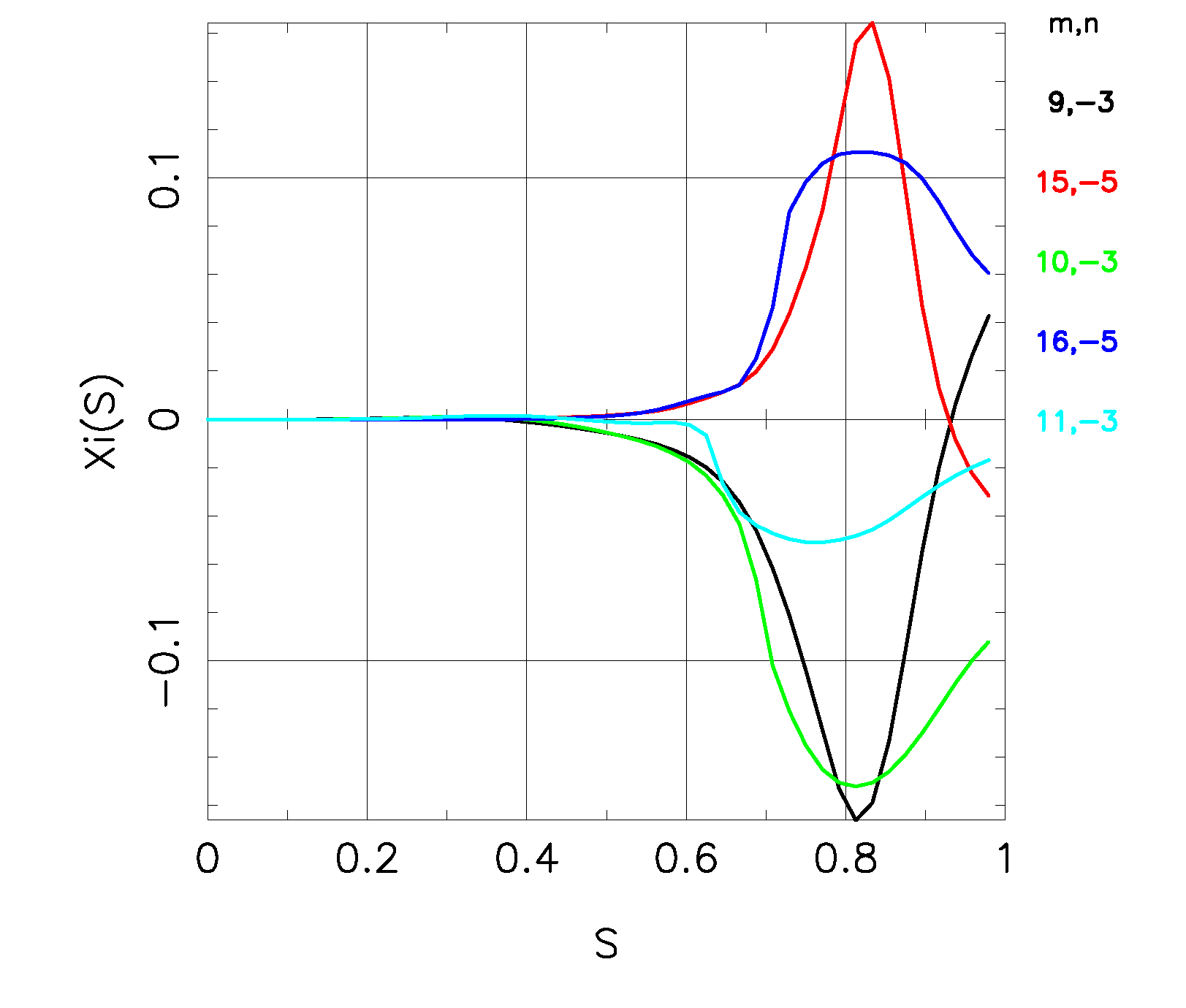} 
		\caption{}
		\label{fig13.b}
	\end{subfigure}
    \caption{Radial structures of Fourier harmonics of external kink mode at $\beta\approx4\%$ for
	(a)the $N=0$ family and (b)the $N=1$ family}
	\label{fig13}
\end{figure}

  While a magnetic well structure suppresses pressure-gradient-driven interchange instabilities, it does not preclude the current-driven kink instabilities. 
To assess these instabilities, we compute the growth rates and mode structures of global external kink modes using the 3D stability code TERPSICHORE\cite{anderson1990methods}. For both CSSC and the optimized configuration at  $\beta=1\%$. 
the computed growth rate eigenvalues are negative, indicating that the ideal global MHD modes are stable. 

  We further investigate the global MHD stability at higher beta values and determine the critical beta.
By maintaining a uniform temperature profile ($T_{e0}=T_{i0}=100eV$) and scaling the density profile with beta, 
we first obtain free-boundary equilibria with self-consistent bootstrap currents  using VMEC  and then evaluate the MHD stability using TERPSICHORE at different beta values. 
\figref{fig12} presents the growth rates of two mode families as a function of beta, 
revealing a critical beta value of approximately $3.2\%$ to $3.5\%$ for MHD stability. 
\figref{fig13} illustrates the radial mode structures of unstable external kink modes at $\beta\approx4\%$, indicating the dominating poloidal components of each mode family, where the x-axis is $S=\psi/\psi_{edge}$. 

\section{Conclusions}
In conclusion, we have obtained optimized finite beta stellarator configurations with four simple coils through optimizing CSSC configuration via coil currents. Our optimization successfully mitigates the finite-beta effects on the effective helical ripple, reducing the ripple level at finite beta to levels comparable to that of CSSC in vacuum ($\beta=0$), especially in edge region. The optimization is achieved by only adjusting the coil currents while keeping the CSSC coil shapes fixed. This is very desirable for an experimental stellarator device where good neoclassical confinement can be maintained at finite plasma beta. 
The detailed properties of the optimized configurations have been presented, including low levels of helical ripple, magnetic well and global MHD stability. This work establishes a novel approach for direct optimization of finite-beta equilibria via external coils, 
demonstrating the existence of optimized finite beta stellarators with simple coils. 
Future work will focus on expanding the optimization framework by incorporating additional degrees of freedom in coil configuration and including more optimization targets. 
This extended approach will enable finer control over plasma performance and further enhance confinement and MHD stability properties of stellarators.
\begin{acknowledgments}
We are indebted to Dr. Steve P. Hirshman for use of VMEC code, Dr. Matt Landreman for use of SFINCS code and Dr. Wilfred A. Cooper for use of TERPSICHORE code. This work was funded by the start-up funding of Zhejiang University for one of the authors (Prof. Guoyong Fu).
\end{acknowledgments}
\nocite{*}
\bibliographystyle{unsrt}
\bibliography{reference.bib}

\end{document}